\documentclass[useAMS,usegraphicx]{mn2e}

\title[Infall of Cold Gas and Escape of Radiation from an Interacting Galaxy]{Filamentary Infall of Cold Gas and Escape of Ly$\alpha$ and Hydrogen Ionizing Radiation from an Interacting High-Redshift Galaxy\thanks{This paper includes data gathered with the 6.5 meter Magellan Telescopes located at Las Campanas Observatory, Chile.}}
\author[Michael Rauch et al.]{Michael Rauch,$^{1}$, George D. Becker,$^{2}$, Martin G. Haehnelt,$^{2}$
\newauthor Jean-Rene Gauthier$^{3}$, Swara Ravindranath$^{4}$, 
Wallace L.W. Sargent$^{5}$\\
$^{1}$Carnegie Observatories, 813 Santa Barbara Street, Pasadena, CA 91101, USA\\
$^{2}$Institute of Astronomy and Kavli Institute for Cosmology, Cambridge University, Madingley Road,  Cambridge CB30HA, UK\\
$^{3}$Dept. of Astronomy and Astrophysics, Kavli Institute for Cosmological Physics, University of Chicago, 5640 S. Ellis Ave,\\
 Chicago IL 60637, USA\\
$^{4}$Inter-University Centre for Astronomy and Astrophysics, Post Bag 4, Ganeshkhind, Pune University Campus, Pune 411 007, India\\
$^{5}$Palomar Observatory, California Institute of Technology, Pasadena, CA 91125, USA}
\begin{document}


\pagerange{\pageref{firstpage}--\pageref{lastpage}} \pubyear{2011}

\maketitle


\label{firstpage}

\begin{abstract}  We present observations of a peculiar  Ly$\alpha$-emitting galaxy  at redshift 3.344, discovered in a deep, blind spectroscopic survey for faint Ly$\alpha$ emitters with the Magellan II telescope in the Hubble Ultra Deep Field (HUDF).  The galaxy exhibits complex Ly$\alpha$ emission, including an extended, asymmetric component that is partially suppressed by damped Ly$\alpha$ absorption, and two spatially elongated, narrow emission features.  Archival HST ACS imaging shows evidence for tidal disruption of the stellar component.  This V=27 galaxy appears
to give us unprecedented insights into two fundamental stages in the formation of structure at high redshift: the inflow of gas into ordinary galaxies, and the escape of ionizing radiation into the intergalactic medium.  Neutral hydrogen, falling in partly in form of a narrow filament, appears to emit fluorescent Ly$\alpha$ photons induced by the stellar ionizing flux escaping from the disturbed galaxy.  The in-falling material may represent primary cold accretion or an interaction-triggered inflow.  The rate of ionizing photons required by the observed Ly$\alpha$ emission  is consistent with the rate of photons produced by the observed stellar population, with roughly 50\% of ionizing photons escaping from the immediate galaxy and encountering the in-falling gas.   We briefly discuss cooling radiation and large scale shocks as additional sources for Ly$\alpha$ and ionizing radiation in high redshift galaxies, but find that stellar radiation is likely to be the dominant source of ionization photons for most faint galaxies.  The observational properties of the galaxy lend support to a picture where galaxy interactions facilitate the escape of both Ly$\alpha$ and ionizing radiation. We argue that galaxies like the present object may be common at high redshift. This galaxy may therefore be a late example of an interacting population of dwarf galaxies contributing significantly to the reionization of the universe.
\end{abstract}


\section{Introduction}


A variety of observations now indicate that hydrogen reionization probably occurred within the redshift range $6 < z < 15$.  Limits on the reionization epoch come from studies of the microwave sky (e.g., Spergel et al. 2003), the Ly$\alpha$ forest seen in absorption against both high-redshift QSOs (Fan et al 2006; Bolton \& Haehnelt 2007; Becker, Rauch \& Sargent 2007, Komatsu et al 2011) and  GRBs (Salvaterra et al 2009; Tanvir et al 2009), and wide-area surveys for high-redshift Ly$\alpha$ emitters (Ouchi et al 2010, Kashikawa et al 2011).  While current constraints remain somewhat weak, further progress in understanding the timing and details of reionization promises to provide unique insights into the nature of the first luminous sources (see, e.g., Ciardi \& Ferrara 2005, Meiksin 2008, and Robertson et al 2010 for reviews). 

 
The nature of the main sources of ionizing photons for the  reionization of hydrogen is a subject of particular contention. The emissivity of QSOs appears to drop too rapidly towards higher redshift for them to contribute significantly (Rauch et al 1997; Bolton \& Haehnelt 2007). There has recently been major progress, however, in identifying possible galactic sources. The redshift limit for the discovery of galaxies has been pushed to $z\sim8$, and possibly  $z\sim 10$, with the new WFC3 camera aboard the Hubble Space Telescope (e.g., Lehnert et al 2010; Bouwens et al 2011).  As the ionizing emissivity inferred for the observed population of  galaxies at $z\sim 6-10$  appears to  fall short of what is needed to reionize the Universe and keep it ionized,  faint galaxies are currently the prime contender for driving hydrogen
reionization (Trenti et al 2010; Bouwens et al 2011).  


The ability of faint galaxies to reionize hydrogen, however, depends critically on the rate of production of ionizing photons within these galaxies, and the ability of those photons to escape into the IGM.  Presently, both the spectral shape of stars/galaxies at these wavelength (e.g. Raiter, Schaerer \& Fosbury 2010) and the escape fractions are highly uncertain.  Escape fractions of ionizing photons are alternately quoted in absolute terms, or as an amount relative to the escape fraction of longer-wavelength ($\sim$1500~\AA) UV continuum photons.  Observed absolute and relative escape fractions appear to increase with increasing redshift and decreasing luminosity  from essentially zero in the local universe (Hurwitz et al 1997; Deharveng et al 2001) to  generally less than 10 percent at redshifts 1 - 3 (e.g., Giallongo et al 2002; Fernandez-Soto, Lanzetta \& Chen 2003; Inoue, Iwata \& Deharveng  2006; Chen, Prochaska \& Gnedin  2007; Siana et al 2007, 2010; Iwata et al 2009; Bridge et al 2010;  Vanzella et al 2010; Nestor et al 2011; Boutsia et al 2011).  Significantly  higher values, however, have been claimed at $z \sim 3$ (Steidel et al 2001; Shapley et al 2006). Observational error may explain some of the discrepancies among the higher redshift studies, but the large variance among the measured escape fractions of individual galaxies could have physical explanations as well.  The escape of ionizing photons could be a transient or recurrent phenomenon,  with individual galaxies being observed in "on" or "off" stages. Variable amounts of dust, and non-isotropic emission of ionizing photons may also be important.
It should also be emphasized that the above observational estimates apply to bright galaxies.  It remains to be seen if the faint galaxies expected to dominate the emissivity of hydrogen ionizing photons at high redshift exhibit relative escape fractions $>50\%$, as apparently required by HI reionization.

Another current impediment to an understanding of the epoch of reionization is the lack of observational constraints on the physical mechanism for the escape of ionizing photons.  For HI ionizing photons to escape galaxies, the medium surrounding the ionizing source (most likely hot stars or AGN) must be optically thin below the Lyman limit. This can be achieved by either a reduced neutral fraction in the surrounding gas, produced by the ionizing radiation field itself, e.g., through the ionization cone of a QSO, or by actual, physical clearings in the gas distribution of the interstellar medium and gaseous halo of the galaxy. The latter may conceivably be produced during interactions with other galaxies or by galactic winds, punching holes into the gas, and tearing open part of the gaseous halo and at least temporarily exposing sources of ionizing photons.

The same processes may release both ionizing (Lyman continuum) radiation into intergalactic space and facilitate the escape of Ly$\alpha$ photons. Unlike the Lyman continuum, Ly$\alpha$ line radiation can escape through scattering by neutral hydrogen in real and frequency space. It does not necessarily require holes or optically thin conditions.   Galactic interactions, through  tidal forces, ram pressure stripping, mass inflow, induced star formation, AGN activity,  compressional heating, or enhanced ionization due to nearby close companions might all serve to reduce the Ly$\alpha$ opacity of the galactic halo and perhaps  open up channels for ionizing radiation. The effects of mergers on the radiative transfer in high redshift galaxies may already have been observed. Cooke et al (2010) have argued that the presence of clustering and impending mergers among galaxies may determine whether a galaxy appears as  a Ly$\alpha$ emitter. A causal connection between the merger rate increasing with redshift and an increase in the escape fraction of ionizing radiation has been suggested by Bridge et al (2010). It clearly would be desirable to be able to directly examine  the escape of Ly$\alpha$ and Lyman continuum radiation in individual interacting galaxies.

\smallskip

A large escape fraction of ionizing photons  during a merger offers another benefit: the possibility that these photons will fluorescently illuminate the  inflowing cold gas from within the galaxies. Inflow of cold ($10^4 K$) gas is believed to be the dominant mode by which low mass galaxies gain most of their baryonic material for star formation  (White \& Rees 1984,  Keres et al 2005), though it has so far eluded direct observational detection.  The anticipated fluxes from cold accretion (e.g., Faucher-Giguere et al 2010; Goerdt et al 2010; Dijkstra \& Loeb 2009) suggest that the prospects of seeing accretion filaments by means of their own Ly$\alpha$ cooling radiation are dim, other than in extremely massive galactic halos, unless an additional source of ionizing photons provides external illumination. Fluorescence of Ly$\alpha$ has indeed been observed several times in the gas associated directly with the QSO environment  (e.g., M\o ller et al. 1998, 2004; Bergeron et al. 1999; Leibundgut \& Robertson 1999; Fynbo et al. 2000; Bunker et al. 2003; Weidinger et al. 2005; Francis \& McDonnell 2006; Hennawi et al 2009) or somewhat further afield from the QSO (e.g., Adelberger 2006; Cantalupo, Lilly \& Porciani 2007). Those fluorescent detections not directly spatially associated with the QSO are likely to mostly come from low mass galaxies, the growth of which should be dominated by cold mode accretion. The number  of intrinsically-produced ionizing photons for even a $z\sim 3$ dwarf galaxy with a mass of a few $\times 10^{10}$ M$_{\odot}$ is sufficient to produce an observed Ly$\alpha$ flux on the order of a few $\times 10^{-18}$erg cm$^{-2}$ s$^{-1}$, which is detectable in a reasonably-deep spectroscopic survey (Rauch et al 2008).  Nearly all the ionizing photons, however, are likely to be absorbed in the interstellar medium or immediate galactic halo, and the properties of the escaping Ly$\alpha$ emission mainly reflect radiative transfer through the galaxy.  For  a galaxy to induce detectable fluorescence {\it in its more distant HI environment, e.g., in gas falling in through cold accretion filaments,} requires (a) that the HI cocoon surrounding the hot stars is breached to let out the ionizing radiation, and (b) that a significant fraction of the ionizing photons  hit the incoming (or otherwise) HI filaments.  
 
\smallskip

In this paper, we discuss observations of a peculiar Ly$\alpha$ emitting galaxy at redshift z=3.344 where the above conditions may apply.  Combining ground-based spectra from the Magellan-Clay telescope and archival imaging from the Hubble Space Telescope, we have detected spatial and spectral emission patterns in this object that suggest a galaxy undergoing a tidal interaction or a merger event. The observational properties of the galaxy may provide us with a (so far) probably unique insight into the escape of Ly$\alpha$ resonance line radiation and ionizing flux at high redshift, while yielding  an observational glimpse into how ordinary high-redshift galaxies accrete gas.

\section[]{Observations}

\subsection[]{Spectroscopic data}

The spectra presented here were obtained in the course of a blind long-slit survey for faint Ly$\alpha$ emission at redshift $z \sim 3$ in the Hubble Ultra Deep Field (HUDF). We have used the LDSS3 spectrograph on the Magellan II Clay telescope at Las Campanas Observatory (LCO) with the VPH Blue grism and a custom ($2"\times 8.3'$) long slit mask. The mask was oriented at a constant position angle of $0\deg$ (i.e., precisely North-South). The orientation was monitored by moving the telescope N-S by several arc-minutes and recording the movements of stars relative to the slit.

A total of 61.4 hours of exposure time was obtained during November 18-23, 2008 and November 11-16, 2009.  Individual exposures were 3000s, and conditions were generally photometric with seeing between 0.45 and 1.0". The exposures were dithered along the slit in intervals of 15 arc-seconds. The spectra were recorded in 1x1 binning on pixels of size $0.189"\times 0.682\AA$ . The $1-\sigma$ detection threshold for extended emission in  a one-square arcsecond aperture amounts to approximately  $9.8\times10^{-20}$ erg cm$^{-2}$s$^{-1}$arcsec$^{-2}$. The slit width-limited spectral resolution for the 2" wide slit was measured to be $\sim$340 km\,s$^{-1}$ (FWHM).  The survey volume is $2056\ h_{70}^{-3}$Mpc$^3$, at mean redshift 3.33. The survey will be described fully, together with other results, in a future paper.

\subsection{HI Ly$\alpha$ emission spectrum}

\begin{figure}
\includegraphics[scale=.32,angle=0,keepaspectratio = true]{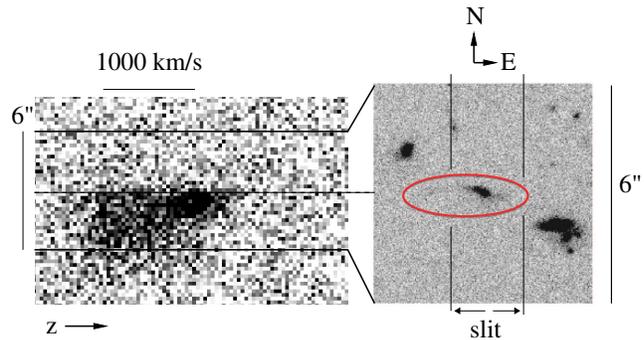}
\caption{Left panel: Part of the two-dimensional spectrum around the Ly$\alpha$ emission line, with the spectral dispersion running horizontally (blue to the left, red to the right), and the spatial direction along the slit vertically (North is up). The section is centered on the wavelength of the compact core of the emission, and on the position of the faint continuum in the spatial direction. The continuum itself is absorbed around the Ly$\alpha$ line regions and thus not visible, but its spatial position is indicated by the thin dashed line. Right panel: F606W ACS image of UDF ACS 07675. The image size is $6"\times6$". Note the change in spatial scale between spectrum and image. The orientation is the same as in the spectrum, i.e., North is up and East is to the right.  Faint emission extends to both sides of the galaxy. The protuberance to the left (West) can be traced over 1.5". The two thin vertical lines indicate the slit jaws.\label{fig1}}
\end{figure}

The object in question was serendipitously discovered as a conspicuously shaped emission line  at an observed wavelength of 5281 \AA\ (fig. \ref{fig1}, left panel; fig. \ref{smoothed}).
An identification as HI Ly$\alpha$ emission at redshift 3.344 is suggested by the drop in the continuum flux blueward of the emission line, by the presence of an apparent damped Ly$\alpha$ trough (fig. \ref{dlatrough}), and by the large extent of the line emission in velocity space and spatially along the slit.

The emission line consists of a sharp, peaked feature with a velocity FWHM of 330 km\,s$^{-1}$ and  a spatial FWHM of 1.5", with a "fan" of emission extending to the blue and to the South (fig. \ref{fig1}; fig. \ref{smoothed}). The compact feature with its red shoulder and drop on the blue side is reminiscent of typical Ly$\alpha$ emitters at high redshift, outside of the Ly$\alpha$ blob luminosity regime (e.g., Rauch et al 2008). We shall refer to this structure as the "red core". In contrast to the compactness of that feature,   the maximum spatial extent of the fan in the slit direction is about 5" ($\sim 37$ kpc proper), if measured down to a flux density level $1.25\times10^{-20}$erg cm$^{-2}$s$^{-1}$\AA$^{-1}$.  The total extent in observed wavelength at that flux density is about 29.5 \AA\ or 1676 km\,s$^{-1}$. The fan appears to have further substructure in the form of two emission "ridges" (see fig. \ref{smoothed}) that are discussed in more detail below.  A faint continuum of approximately 27.07 $m_{AB}$ in the V band is detected in the two-dimensional spectrum, offset by 0.8" to the North of the peak of the Ly$\alpha$ emission (fig. \ref{dlatrough}). The continuum shows evidence  of a damped Ly$\alpha$ absorption trough surrounding the wavelength position of the emission feature. The total flux in the Ly$\alpha$ emission line is F=$2.45\times10^{-17}$erg cm$^{-2}$s$^{-1}$, of which only $5.11\times10^{-18}$erg cm$^{-2}$s$^{-1}$ are emitted from the compact red core of the line. Thus, about 80\% of the total Ly$\alpha$ emerges in the extended blue fan (fig. \ref{smoothed}).  The restframe equivalent width of the entire complex is $(79 \pm 5)$~\AA .

\subsection{Broad band continuum imaging of the underlying galaxy}

The closest continuum object in the HUDF, 0.8" from the absolute spatial position predicted on the basis of the position of the emitter along the slit, is the galaxy GOODS-CDFS-MUSIC 11517 (a.k.a. UDF ACS 07675; 03:32:38.815 -27:46:14.34 (2000)). Two published photometric redshifts, 3.556 (Coe et al 2006) and 3.38 (Ryan et al 2007), support the reality of the object as a $z=3.344$ Ly$\alpha$ emitter.  The published V band magnitude, $m_{AB}$=26.976 (Thompson 2003)  agrees well with the continuum magnitude of our spectrum (see above), confirming the impression gleaned from an inspection of co-added through-slit images, that virtually all of the continuum flux went down the slit (but not necessarily all of the Ly$\alpha$ flux).

A V band (F606W) HST ACS image (Beckwith et al 2006) is in shown in the RHS panel of fig.\ref{fig1}. The galaxy looks highly disturbed in the continuum image, with a 1.5" (11 kpc proper) long 'spur' to the W, and another shorter protuberance to the SE. The star formation rate, estimated based on the F606W UV continuum according to the prescription of Madau,  Pozzetti, \& Dickinson (1998), and uncorrected for extinction, is SFR = 1.7 M$_{\odot}$yr$^{-1}$.

\subsection{He II 1640 \AA\ emission spectrum}

Additional low resolution spectroscopy for the galaxy is available from the archive of the GRAPES survey (Pirzkal et al 2004).  The observed wavelength range covers among other things the redshifted HeII 1640 \AA\ emission. Adding the spectra from all orientations, we obtain a $5-\sigma$ detection of HeII 1640 \AA\ emission, at 7106 \AA , with a flux of $(2.4\pm 0.5)\times10^{-18}$erg s$^{-1}$cm$^{-3}$. This is a factor  ten fainter than the  HI Ly$\alpha$ flux.  HeII 1640 at this redshift may have a number of different origins, as it is  generally indicating the presence of moderately hot gas. It occurs in winds from Wolf-Rayet stars (e.g., Schaerer \& Vacca 1998), is predicted to be present in Population III stars (Schaerer 2003), and is found in the halos of radiogalaxies (e.g., Villar-Martin et al 2003). It may also occur as cooling radiation during structure formation (Yang et al 2006; Scarlata et al 2009). However, the intensity of cooling radiation in a quiescent galactic halo is a strong function of the halo mass (e.g., Dijkstra et al 2006), and for the likely low mass of the underlying galactic halo cooling radiation would be too faint to be detected in our spectrum  in HeII 1640.

\section[]{Detailed analysis of the L\lowercase{y}$\alpha$ emission line structure}

As mentioned earlier, the basic features of the Ly$\alpha$ region line profile can be divided into (1) a relatively sharp red core, (2) a diffuse blue fan in emission, and (3) a damped Ly$\alpha$ system in absorption.

\subsection[]{Foreground DLA absorption}

\begin{figure}
\includegraphics[scale=.6,angle=0,keepaspectratio = true]{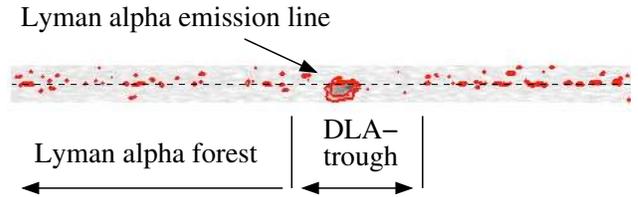}
\caption{Smoothed 2-D LDSS3 spectrum of a 640 \AA\ long  wavelength strip surrounding the Ly$\alpha$ emission region, emphasizing the features of the galactic continuum. The spectrum is 10" wide, the orientation is the same as in the previous figure. The position of the continuum is shown by the thin dashed line. A gap in the continuum emission surrounds the Lyman emission line, possibly asymmetric in wavelength with respect to the line position (the precise extent of the gap is hard to tell given the low signal-to-noise ratio in the continuum). Interpreting the gap as damped Ly$\alpha$ absorption trough also explains the sharp cutoff at the northern (top) end of the Ly$\alpha$ emission line region as absorption by the foreground galaxy of the background emission region. 
 \label{dlatrough}}
\end{figure}

\begin{figure}
\includegraphics[scale=.31,angle=0,keepaspectratio = true]{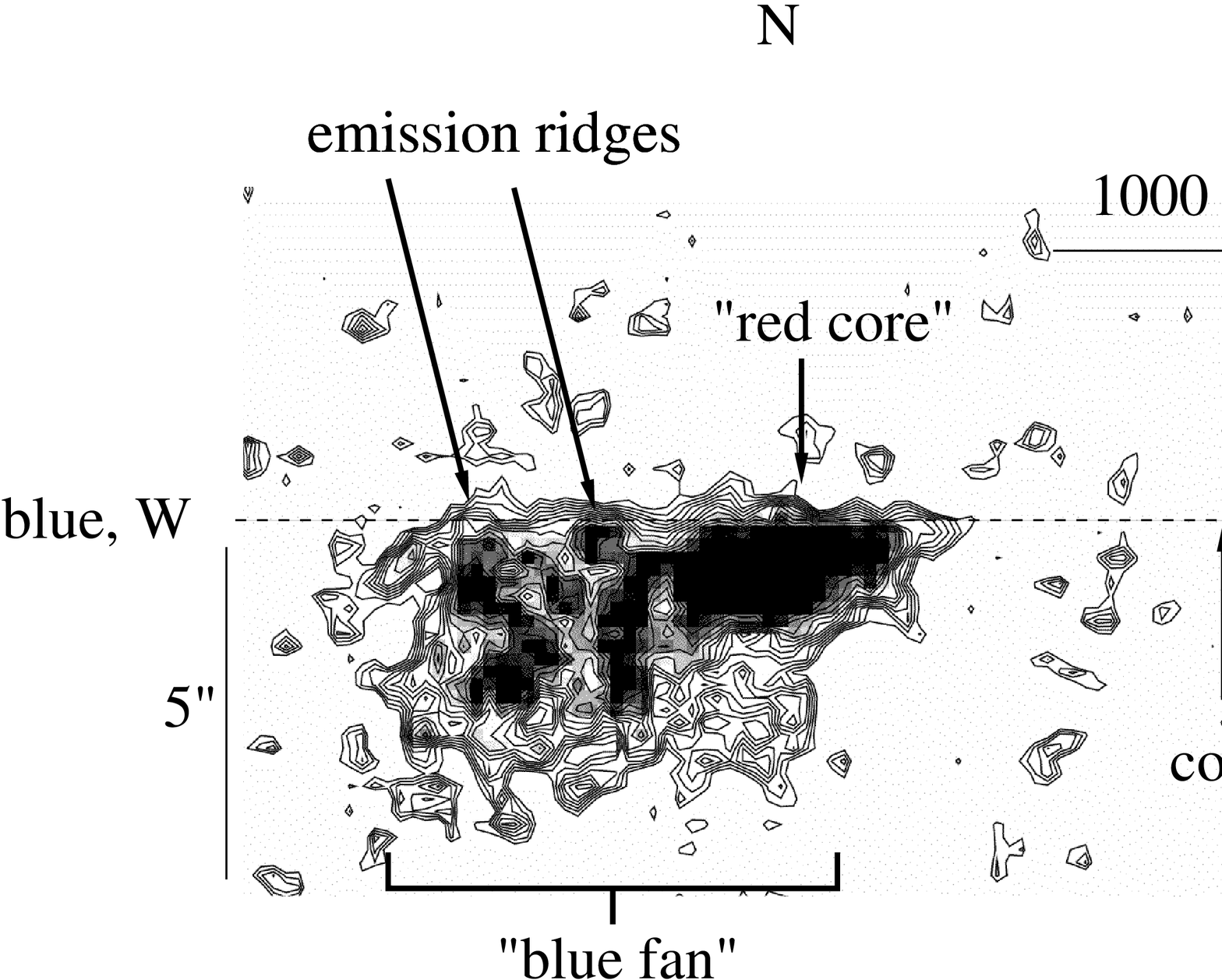}
\caption{ The spectrum has been smoothed with a 3x3 pixel ($0.57"\times118$km s$^{-1}$)  boxcar filter.  The intensity is denoted by a combined contour/grey level plot. The outermost contiguous contour corresponds to a flux density $1.25\times10^{-20}$erg cm$^{-2}$s$^{-1}$\AA$^{-1}$, the average flux density in the emission ridges is $5\times10^{-20}$erg cm$^{-2}$s$^{-1}$\AA$^{-1}$.  For comparison, the $1-\sigma$ noise per pixel prior to smoothing is  $2.7\times10^{-20}$erg cm$^{-2}$s$^{-1}$\AA$^{-1}$  \label{smoothed}}
\end{figure}

An absorption trough extends in wavelength across the entire emission region (fig.\ref{dlatrough}).  Although damping wings cannot be directly detected due to the low signal-to-noise ratio of the continuum, the large velocity width of the absorption strongly suggests that this is a damped Ly$\alpha$ absorber (DLA).  The same gas cloud absorbing the galactic continuum also appears to cause the sharp drop in flux of the emission feature northward  of the position of the continuum trace (fig. \ref{smoothed}).  Obviously, {\it the DLA host galaxy lies between the observer and the gas clouds emitting Ly$\alpha$}.  Because of the low signal-to-noise ratio of the continuum and the presence of the emission line complex, the exact extent of the DLA trough is hard to discern, especially on the blue side. Extrapolating the partial equivalent width from the visible red wing of the trough to the position of the center of the emission line, and assuming that the trough continues by an equal extent to the blue of the emission line,  a crude estimate of about 40 \AA\ is obtained for the equivalent width in the galaxy's rest frame. corresponding to an HI column density of N$_{HI}\sim 3\times10^{21}$ cm$^{-2}$.  This value is on the high side  for lines-of-sight going through random regions of DLA absorption, but not unusual for DLAs caused by the host galaxies of gamma-ray bursts (GRBs) (e.g., Vreeswijk et al 2005; Chen et al 2007), which are known to be associated with the central, star-forming regions of relatively low mass galaxies, like the present object. The distance along the slit between the continuum trace of the galaxy and the peak (of the visible part) of the red core is about 0.8" or 6.1 kpc. The projected separation along the slit between the continuum and the sharp cutoff of the Ly$\alpha$ line is less than about 3 kpc (i.e., within the spatial resolution along the slit), suggesting that the DLA absorbing cross-section has a radius of similar value. This is  consistent with earlier suggestions (e.g., Tyson 1988; Fynbo et al 1999; Haehnelt et al 2000; Rauch et al 2008; Rauch \& Haehnelt 2011) that DLA hosts are predominantly small, low mass galaxies. 

\subsection[]{Red core  vs. blue fan}

The wavelength separation between red and blue features may partly represent actual  motion of the gas in the fan relative to the core, and partly wavelength drift caused by resonance line scattering. If the Ly$\alpha$ emission from the core were the red wing of a (partly suppressed) double-humped structure, its wavelength would not indicate the systemic velocity of the galaxy, and the actual kinematic velocity extent could be much smaller.  An apparent velocity difference as observed between the red core and the center of the blue fan ($\Delta v \sim 670$km s$^{-1}$) can indeed be entirely produced between the two humps of the emission line profile emerging from scattering in a static gas cloud with a column density  of  \begin{eqnarray} N_{\rm HI}=5.3\times10^{20}\left(\frac{T}{10^4{\ \rm K}}\right)^{-1/2}\left(\frac{\Delta v}{670 {\ \rm km\,s}^{-1}}\right)^3 {\rm cm}^{-2}\label{slab} \end{eqnarray}  (e.g., Dijkstra et al 2006; Hansen \& Oh 2006; and references therein).

However, the relatively stronger intensity of the blue vs. red emission suggests that we are not dealing here with a static configuration.  The dominance of the blue over the red part of a Ly$\alpha$ profile as observed here does instead indicate infall (e.g., Zheng \& Miralda-Escud\'e 2002; Dijkstra et al 2006; Verhamme et al 2006), but the spatial asymmetry of the observed profile, with the much larger spatial extent for the blue fan, and its tilt to the South relative to the red core  makes it unlikely that the blue fan is a direct counterpart to the red core, caused by resonant scattering of Ly$\alpha$ photons all originating in the same location.  Assuming that the Ly$\alpha$ emission is produced by photoionization from a single galactic source (e.g., stars or an AGN) the lack of coherence between the blue fan and the red core may instead be due to the fact that both are physically distinct bodies of gas hit independently by ionizing UV radiation. 

\subsection[]{Substructure of the blue fan}

The fan appears to consist of mostly amorphous flux, superposed on which appear two noisy, ragged "ridges" of emission (fig. \ref{smoothed}), shifted to the blue of the compact emission core by about 670 km\,s$^{-1}$.  The velocity separation between the ridges is about 320 km\,s$^{-1}$.  It is difficult to estimate the flux in the ridges but it appears to be not more than about 25\% of the total flux in the fan. 

While the ridges are clearly significant assuming purely statistical errors, they appear considerably narrower ($\sim 160$ km\,s$^{-1}$ FWHM) than the spectral resolution  for an extended object observed  through a  2" wide slit (340 km s$^{-1}$ at the observed wavelength), cautioning us that they could be artifacts. The structure does appear to persist (albeit at a noisier level) if the data obtained in 2008 and 2009 are considered separately. We have examined the position of the emitter relative to the bridges in the spectrograph mask that could have produced reflections along the slit but found that the object always was at least 5" away, and generally much further from such features. Thus, there is no obvious reason to think that the ridges are not intrinsic to the emission pattern.

The narrowness of the ridges is naturally explained, however, if the emitting gas does not fill the entire slit. For example, if the emission were dominated by a Ly$\alpha$-emitting hotspot or filament with an observed thickness equal to that of the seeing disk (the effective seeing after co-adding all frames was $\sim 0.75"$), then the seeing-limited velocity resolution would be $\sim 125$ km\,s$^{-1}$, and not the 340 km\,s$^{-1}$ expected for an extended object. 

If the ridges are real, then they may simply be due to individual sources, e.g., multiple galaxies or HII regions, or the shredded bits of the stars and interstellar medium of the disturbed galaxy,  each contributing their own Ly$\alpha$ emission line at wavelength positions reflecting their relative motions.  The peculiar appearance of the the two-ridge pattern, however, suggests that it instead may be caused by resonance scattering of Ly$\alpha$, rather than having a purely kinematic origin.  

The two ridges appear more consistent with the double-humped Ly$\alpha$ emission line expected from a static cloud than does the overall emission complex as a whole. For gas at a temperature of 10,000 K, the velocity separation between the emission ridges implies a column density of N$_{\rm HI}$=$5.8\times10^{19}$cm$^{-2}$ (eqn. \ref{slab}), which is characteristic of the so-called "sub-damped" Ly$\alpha$ systems. The similar intensities of the emission ridges implies that the gas giving rise to these features may not have a significant internal velocity gradient, which would tend to enhance one peak relative to the other.  Thus, the particular spatial region of the gas responsible for the emission ridges may indeed be a kinematically quiescent, optically thick filament of gas that functions as a simple, fluorescent 'mirror', 'reflecting' the incoming UV radiation from the source in the form of Ly$\alpha$ toward the observer. The justification for calling this a 'filament' rests on the projected aspect ratio in the direction along the slit (total length/thickness $\ge 3$) and the inferred quiescence in velocity space.  For a spatially resolved narrow filament with a diameter D and column density $N_{\rm HI}$, we can estimate the neutral hydrogen density as 
\begin{eqnarray}
n_{\rm HI} = 2.5\times 10^{-3}{\rm cm}^{-3}\left(\frac{N_{\rm HI}}{5.8\times 10^{19}\ {\rm cm}^{-2}}\right)\left(\frac{D}{1"}\right)^{-1}.
\end{eqnarray}
This corresponds to an overdensity of about 150 with respect to the mean density at z=3.344, assuming that the hydrogen in the filament is predominantly neutral. The total HI mass depends on the length of the filament, $L$, which we see only in projection in the plane of the sky: 
\begin{eqnarray}
m_{\rm HI} > 6\times 10^7 M_{\odot}\left(\frac{N_{\rm HI}}{5.8\times 10^{19}\ {\rm cm}^{-2}}\right)\left(\frac{D}{1"}\right)\left(\frac{L}{2.7"}\right).
\end{eqnarray}

Although the signal-to-noise ratio in these features is modest, the ridges appear to be tilted by roughly 160 km\,s$^{-1}$ over a spatial distance of about 2.7".  Such a tilt could be caused by rotation. Alternatively, if the ridges are indeed caused by a narrow filament not filling the slit, a geometric tilt in position space relative to the slit can cause different parts of the filament to land at different angular position in the direction across the slit. For the current instrumental configuration, a shift of 1" to the East would lead to an apparent velocity shift of 188 km\,s$^{-1}$ to the red, for example. A third possibility is that such filament could be orientated away from the observer, perpendicular to the plane of the sky, and undergoing accelerated infall. This is consistent with the increasing blueshift of the ridges when approaching the continuum trace of the galaxy from the South.  Such acceleration would be expected for any infall into the gravitational potential well of a galaxy, and, presumably, also be one of the likely observational signatures for cold accretion flows, could they be seen in Ly$\alpha$ emission.

\subsection[]{A plausible model}

\begin{figure}
\includegraphics[scale=.15,angle=0,keepaspectratio = true]{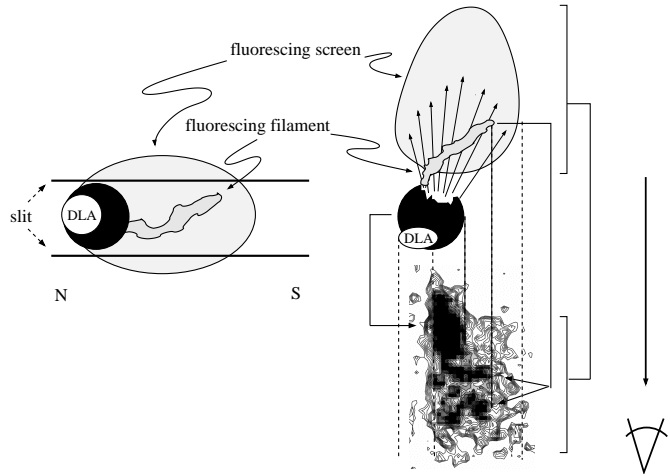}
\caption{A plausible configuration that appears to satisfy the observational constraints. The left side shows a view of the object as it could appear through the slit (for display purposes the slit is rotated by $90\deg$ relative to the previous figures). The spatial shapes are largely arbitrary.  In this picture, the galaxy produces a faint continuum, the red core of the Ly$\alpha$ emission, and the DLA absorption. A narrow filament extends over several arcseconds but does not fill the slit. A larger background cloud fills the slit over 5".  The angle of the filament relative to the slit in the plane of the sky may introduce the observed tilt in wavelength space; alternatively, the filament could be tilted along our line of sight, and we could be seeing accelerated infall toward the galaxy. The damped Ly$\alpha$ system absorbs everything that is physically behind it, including the continuum of the galaxy and part of all Ly$\alpha$ emission features to the north of the galaxy continuum.   The right hand side shows the proposed topography from above, and how it relates to the spectral features, with the direction to the observer shown on the right. The Ly$\alpha$ emitting galaxy halo, as far as it is not absorbed by the smaller DLA region and the intergalactic medium, produces the red core. The filament and the larger screen (which may or may not be physically connected to the galaxy) are exposed to ionizing radiation escaping from  the disturbed halo. They re-radiate part of it as Ly$\alpha$, in the form of a narrow double humped profile, or a turbulently broadened and spatially extended fan, respectively. Note that, to produce the blue-shifted spectral features, both the filament and the screen have to fall toward the galaxy from the backside. The fact, that the largest visible extent of the blue fan along the slit direction occurs at the blue end of the emission suggested that the region closest to the galaxy (that is exposed to the largest UV flux and thus can be seen out to the largest distance) falls in with the largest velocity. The slight tilt of the emission ridges to the blue as approaching the galaxy could be explained by the same pattern, increasing infall velocity with decreasing radius.\label{scheme}}
\end{figure}

The distribution of gas around interacting galaxies should be highly complex, and may include bound interstellar material, extended debris from tidal stripping, and the local network of intergalactic filaments.  While it may not be possible to uniquely explain the observed characteristics of this system, however, we can still identify a relatively simple scenario that may potentially deliver insight into a number of processes affecting high-redshift galaxies.  Our model is summarized in figure (\ref{scheme}).  As a starting point, the DLA is likely to be related to the HI content of the 27 mag galaxy. As the DLA absorbs not only the galactic continuum but also the northern edge of the Ly$\alpha$ emission complex, it must be in front of both. The red core emission is sufficiently close spatially to the continuum trace of the galaxy and the DLA absorption to arise in the same object, whereas the fan with its blue shift is moving toward that galaxy on the backside. The fact that the largest spatial extent of the emission along the slit  occurs at the bluest wavelengths fits into this picture if the inflow accelerates with decreasing distance from the galaxy.  The largest velocities would occur closest to the galaxy, where the flux of the ionizing radiation is highest, allowing the illuminated gas to be seen further out in the direction along the slit.    The sharp emission ridges would also arise from in-falling but rather more quiescent and more spatially compact material, possibly a cold flow filament or tidal debris. The tilt of the ridges in velocity space toward the galaxy  (bluer, where closer to the  galaxy's continuum trace, see above) again suggests accelerated infall toward the galaxy. A similar pattern occurs in some of the faint emitters in Rauch et al (2008), (see the discussion in their section 6.3.3).

The asymmetric Ly$\alpha$ emission may be explained if we assume that the gaseous halo has been split open in the direction away from the observer. As suggested in fig. \ref{scheme}, the ionizing radiation would escape from the galaxy, inducing Ly$\alpha$ fluorescence in the infalling gas clouds. Alternatively, the apparent tidal stellar features seen in the broad band image may have been stripped of neutral hydrogen through the interaction, providing a "naked" source of ionizing photons. One reason for believing that  the Ly$\alpha$ emission results from ionizing radiation hitting the screen and filament, rather than from a transfer of Ly$\alpha$ photons propagating from the innermost parts of the galaxy, is  the simple, static velocity structure of the emission ridges. Another reason is the nearly constant Ly$\alpha$ surface brightness in the blue fan up to several arcseconds (tens of kpc) away from the galaxy. This is consistent with the fluorescing surface of an HI screen, but unlike any standard galaxy Ly$\alpha$ surface brightness profiles observed or predicted: For more common "standard" Ly$\alpha$ emitters, Rauch et al (2008) find a rapid drop of the Ly$\alpha$ surface brightness over a few arcseconds from the center, a result well understood (Dijkstra et al 2006, Barnes \& Haehnelt 2010, Zheng et al 2010) if the Ly$\alpha$ propagates outward from a central source of ionization through an optically thick and inviolate HI halo.

\subsection[]{The relation between disrupted L\lowercase{y}$\alpha$ halos and mergers}

If the disruption of a gaseous halo is a consequence of galaxy interactions or mergers, we would expect such peculiar Ly$\alpha$ emission as seen here to be roughly as common as the mergers themselves. How plausible is a serendipitous discovery of a merging event at redshift 3.3 in the present survey ?

The duty cycle for major/significant  galaxy mergers increases rapidly  towards higher   redshift,  along with the  decrease of the characteristic mass $M_{*}$  of  the dark matter haloes hosting the  galaxies (see e.g. Yang et al. 2011 for a recent quantitative assessment of the merger rates of DM haloes). The exact duty cycle  of galaxies for undergoing (major) mergers will obviously depend on the  mass ratio of the merger and the exact definition of the end and start  of the merger, but the corresponding time interval  should be on the order of a Giga-year.  The duty cycle should thus increase  to well above 10\% at high redshift and may be even larger in observed flux-limited samples of high-redshift galaxies, which are expected to favor compact star-forming or -bursting  galaxies.   Thus, our finding of one such object  out of a total of roughly two dozen Ly$\alpha$ emitters (the exact number will have to await the completion of the analysis of the entire field) is expected, and we  have no reason to believe that this is an unusual incidence. Note that, with a duty cycle that large, and escape fractions on the order of 50\%,   mergers of (small) high-redshift galaxies with low dust content may make a significant contribution  to the emissivity of hydrogen ionizing photons.



\section[]{Possible Sources of Ionization}

We have presented a scenario in which the peculiar Ly$\alpha$ features in this system arise from a multi-component distribution of gas around a galaxy that has recently experienced a significant merger or interaction.  We now turn to discuss the possible mechanisms which may be powering the Ly$\alpha$ emission itself.

\subsection[]{Stellar photoionization}

Photoionization by a stellar Lyman continuum appears to be the most likely driving force  behind most  Ly$\alpha$ emitters (Rauch et al 2008).  If both the red core and the blue fan are irradiated by ionizing radiation from the same central source, we have to reconcile the observed fluxes in these features with each other and with the total amount of ionizing photons from that source. As argued earlier, we assume that part of the ionizing radiation is absorbed in the galaxy, leading to a red core of Ly$\alpha$ emission, with a Ly$\alpha$ escape fraction of $f_{\rm core}^{ly\alpha}$ . The rest of the ionizing radiation escapes the galaxy with escape fraction $f^{ll}$ but then gets absorbed partly (with fraction $f_{\rm cov}^{ly\alpha}$) by the external, in-falling gas. The in-falling clouds respond with Ly$\alpha$ fluorescence detected as the blue fan.

The rate of production of ionizing photons, $\dot{N}_{core}^{ion}$ required to explain the observed Ly$\alpha$ flux in the red core ($5.11\times10^{-18}$ erg cm$^{-2}$s$^{-1}$), assuming a ratio of 2/3 Ly$\alpha$ photons per ionizing
photon (e.g., Gould \& Weinberg 1996),  is 
\begin{eqnarray} 
\dot{N}_{\rm core}^{\rm ion} = \frac{3}{2} \frac{F_{ly\alpha}}{h\nu_{ly\alpha}} 4\pi D_L^2\left({1 - f^{ll}}\right)^{-1}\left({f_{\rm core}^{ly\alpha}}\right)^{-1}=\ \ \ \ \ \ \ \ \    \nonumber \\
9.42\times10^{52} \left(\frac{F_{ly\alpha}}{5.11\times10^{-18}}\right) \left({1 - f^{ll}}\right)^{-1}\left(\frac{f_{\rm core}^{ly\alpha}}{0.5}\right)^{-1} {\mathrm s}^{-1},\label{coredot}
\end{eqnarray}
and depends on the Ly$\alpha$ flux $F_{ly\alpha}$, the energy of the Ly$\alpha$ photon $h\nu_{ly\alpha}$, and the luminosity distance $D_L$.  Assuming the usual spatial symmetry of compact Ly$\alpha$ emission, and accounting for the absorption of the northern half of the red core by the DLA, we have adopted a fiducial value of 0.5 for  $f_{\rm core}^{ly\alpha}$.

Similarly, the photon production rate causing the observed flux in the blue fan ($1.94\times10^{-17}$erg cm$^{-2}$s$^{-1}$)  can be written as
\begin{eqnarray} 
\dot{N}_{\rm fan}^{\rm ion} = \frac{3}{2}  \frac{F_{ly\alpha}}{h\nu_{ly\alpha}}   4\pi D_L^2
\left(f^{ll}\right)^{-1}
\left(f_{\rm cov}^{ly\alpha}\right)^{-1}
\left(f_{\rm geo}^{ly\alpha}\right)^{-1} {\mathrm s}^{-1}    =\nonumber \\
9.32\times10^{52} \left(\frac{F_{ly\alpha}}{1.94\times10^{-17}}\right)\times \ \ \ \ \ \ \ \ \ \ \ \ \ \ \ \ \ \ \ \ \nonumber\\
\times \left(f^{ll}\right)^{-1}
\left(f_{\rm cov}^{ly\alpha}\right)^{-1}
\left(\frac{f_{\rm geo}^{ly\alpha}}{2}\right)^{-1} {\mathrm s}^{-1}.\label{screendot}
\end{eqnarray}
Here, $f_{\rm cov}^{ly\alpha}$ is the covering factor of the blue fan, which takes into account the fact that the HI screen producing the fan emission may not trap all of the ionizing radiation. Only a fraction $f_{\rm cov}^{ly\alpha}$ of it will be turned into Ly$\alpha$ emission. We take it to be unity, assuming that all ionizing photons ultimately hit the infalling nearby HI and produce Ly$\alpha$.  If the fan emission is indeed fluorescent we need to correct the way the luminosity is computed from the flux, adjusting for the fact that the radiation only comes out over $2\pi$ instead of $4\pi$. Without knowledge of the orientation of the screen we can only make a basic correction. By introducing a geometric factor $f_{\rm geo}^{ly\alpha}$ with fiducial value 2 we take the ratio between the {\it observed} Ly$\alpha$ flux and the ionizing radiation impinging on the fan to be twice the value for isotropic emission.

Requiring that the same source of ionizing photons produce the core and fan emission according to our assumption, we then equate relations (\ref{coredot}) and (\ref{screendot}), and obtain an escape fraction for ionizing photons (photons that escape the galaxy at least as far as the blue screen) of $f^{ll}\sim 0.5$, which seems a reasonable for the picture suggested here -  it means the backside of the galaxy is essentially fully exposed.  This value for $f^{ll}$ is actually a lower limit, as it assumes that half of the red core flux is lost to the DLA, all of the escaping ionizing radiation hits the blue screen, and all of the Ly$\alpha$ gets reflect into the hemisphere facing the backside of the galaxy and the observer. Alternatively, assuming that only half of the ionizing photons get trapped by the screen raises the escape fraction to 0.67, as does assuming that the red core Ly$\alpha$ emission is unabsorbed, but there is no obvious way of reducing the escape fraction for ionizing photons to less than 50\%.

\smallskip 
For comparison, the number of ionizing photons predicted to be produced by the stellar population is
\begin{eqnarray} 
\dot{N}_{*}^{\rm ion} =\int_{\nu_{ll}}^{\infty} \frac{L_{ll}}{h\nu}\left(\frac{\nu}{\nu_{ll}}\right)^{-\alpha_s} d\nu =
 \frac{L_{1500}}{18 h} \left(\frac{\alpha_s}{3}\right)^{-1}= \nonumber \\
1.2\times10^{53}\left(\frac{\alpha_s}{3}\right)^{-1}\left(\frac{L_{1500}}{1.42\times 10^{28}{\rm erg\ s}^{-1}{\rm Hz}^{-1}}\right)  {\rm s}^{-1},\label{stardot}
\end{eqnarray}
where the ionizing yield of the galaxy has been parametrized following Madau, Haardt \& Rees (1999). We assume a drop in the continuum of $L(1500\AA)/L(900\AA)$=6, and a continuum slope of $\alpha_s$ =3 below 912\AA , where $L\propto \nu^{-\alpha_s}$. Comparing the stellar production rate (\ref{stardot}) with the rates derived from the observations of Ly$\alpha$, we find from eqn. (\ref{screendot}), for an escape fraction $f^{ll}$=0.5, that the observed flux requires a value of $\dot{N}_{*}^{\rm ion}=1.86\times10^{53}$s$^{-1}$, which is higher by only a factor of 1.55 than the predicted rate based on the stellar continuum model. Recent estimates by Haardt \& Madau (2011) increase the galactic yield for ionizing photons by about 1/3, which would bring the two rates in even closer agreement. In any case, a minor adjustment of the stellar model, increasing $\alpha_s^{-1} L_{1500}$ by factors 1.2 - 1.55, allows us to conclude that the Ly$\alpha$ emission can be powered entirely by the stars we infer from the galactic continuum redward of Ly$\alpha$ emission.

We note that not all of the Ly$\alpha$ emission may have gone through the 2" wide long slit.  We also do not know the details of the spatial gas distribution, and so the agreement between the production rate of ionizing photons required by the observations and the actual stellar production rate based on the continuum luminosity could be fortuitous. It may therefore be worth considering whether the Ly$\alpha$ emission can be produced by other astrophysical mechanisms.

\subsection[]{Cooling radiation}

The possibility of Ly$\alpha$ from the cooling of gas during gravitational contraction has been  discussed mainly in connection with the so-called Ly$\alpha$ blobs (e.g., Fardal et al 2001, Dikstra \& Loeb 2009, Faucher-Gigu\`ere et al 2010).  Using Faucher-Gigu\`ere et al's analytic relation for the star formation-induced Ly$\alpha$ emission as a function  of halo mass (their equation A10), and assuming for a moment that all measured Ly$\alpha$ luminosity ($1.23\times10^{42}$ erg s$^{-1}$)  were of stellar origin for our object, we would estimate a total mass for our galaxy  of $2.2\times10^{10}$M$_{\odot}$. For this value the expected luminosity of the cooling Ly$\alpha$ would be a factor $\sim 23$ lower than the stellar one. As the ratio between cooling and stellar Ly$\alpha$, $L_{Ly\alpha}^{\mathrm cool}$/$L_{Ly\alpha}^{\mathrm SF}$ rises with total mass only as $M^{0.4}$, the cooling radiation from cold accretion becomes dominant only for much more massive halos, outside of the mass range of typical Ly$\alpha$ emitters.

\subsection[]{Photoionization by AGN}

Our current data cannot exclude that the emitter is  photoionized by an AGN. The large velocity extent of the emission, however, makes it unlikely that the object represents a Str\"{o}mgren sphere or the ionization cone of an AGN.  The velocity width, interpreted as fluorescence of gas in the Hubble flow, would require a line-of-sight extent of several Mpc, when at the same time the lateral diameter of the structure is only about 40 kpc.  The whole structure, therefore, would correspond to a 'needle' of ionization rather than a cone or sphere. The apparent backside inflow of Ly$\alpha$ emitting material also does not suggest an AGN, unless a second object interacting with the foreground galaxy is hosting one. There is also none of the evidence usually associated with  the presence of an AGN;  the Ly$\alpha$ red core emission line has a velocity width of only 330 km\,s$^{-1}$, and  the galaxy is not detected by either the VLA survey of the CDFS (Kellerman et al 2008) or the Chandra 2 Megasec exposure of the CDFS (Luo et al 2008).

\subsection[]{Ionization by fast radiative shocks}

When galaxies interact, the collisions of gas clouds may result in shocks. Part of the energy  is radiated away in the form of ionizing UV photons and UV line radiation, including detectable Ly$\alpha$  and HeII 1640. The hard UV spectrum produced can lead to the photoionization of the precursor (e.g., Dopita and Sutherland 1995) and possibly escape from the immediate shock zone and perhaps from the galactic halo(s), contributing to the general UV background. Recent findings of spatial offsets between the HI ionizing and non-ionizing continuum emission of starforming galaxies  (Iwata et al 2009), and ratios between ionizing and non-ionizing flux that are uncomfortably large if to be explained solely by standard stellar spectrosynthesis modeling (Nestor et al 2011) suggest that there could be additional galactic contributions to the ionizing background, unrelated to stellar or AGN sources of photoionization.

To understand whether galactic scale shocks can play role in the appearance of our Ly$\alpha$ emitter, we can compare the observed Ly$\alpha$ luminosity, L$_{{\rm Ly}\alpha}$=$2.46\times10^{42}$erg s$^{-1}$, and ratio L(HeII 1640)/L(HI Ly$\alpha$)$\sim 0.1$ to plausible values for shocks arising in galactic collisions, using model output from the MAPPINGS library (Allen et al 2008). Obviously, the actual physical parameters of a galactic scale shock would be highly uncertain. 

The details of this estimate are given in Appendix A.  We find that it is well possible that a massive large scale shock, occurring over the entire area (580 kpc$^2$) of the observed Ly$\alpha$ emitting region with shock velocities at least as high as 500 km\,s$^{-1}$ could produce Ly$\alpha$ and HeII 1640  emission as observed, and create an ionizing radiation field with a strength (at the  Lyman limit) sufficient to mimic a 10-20 percent escape fraction for a 27 magnitude galaxy like the one discussed here.  We note, however, that shocks of cold gas over such large areas, and with shock velocities on the order of 500 km\,s$^{-1}$ are unlikely to be gravitational in origin, and one would probably have to rely on stellar/galactic winds, jets or other non-gravitational gas flows. 


Galactic winds of the type supposed to be active in bright Lyman break galaxies are unlikely to be relevant in the present case because of the small star formation rate inferred for our galaxy.  In the few cases where  spectra of comparable depth of the Ly$\alpha$ emission from Lyman break galaxies have been obtained (e.g., Rauch et al 2008; Rauch et al in prep.), the Ly$\alpha$ line generally appears spatially symmetric, and does not show evidence of material disturbed out to tens of kpc, as in the present case, or of dominant blue emission. 

Summarizing, a contribution of ionizing radiation and Ly$\alpha$ emission from shocks is plausible, given the interacting nature of the galaxy, but it is unlikely that shocks will play a dominant role.

\medskip

\section[]{Conclusions}

We have detected a faint star-forming galaxy at $z = 3.344$ with highly peculiar Ly$\alpha$ features as part of a deep, blind spectroscopic survey.  The Ly$\alpha$ emission has multiple components, including a compact, red core similar to those found in other Ly$\alpha$-emitting galaxies, an extended and diffuse blue fan, and two spectrally narrow, spatially elongated ridges.  Part of both the compact and extended emission appears to be blocked by an associated DLA.  Additional insight into this system is provided by archival HST/ACS imaging, which shows tidal tails suggestive of a recent or ongoing interaction.

The spatial and kinematic distribution of gas around an interacting galaxy is likely to be complex, and more than one configuration may plausibly explain the observed features of this system.  Nevertheless, this object provides a unique test case for studying a number of processes affecting high-redshift galaxies.  We have described a simple scenario that highlights, in particular, the importance of such objects for understanding both the fueling of star-forming galaxies with cold gas, and the escape of ionizing and Ly$\alpha$ radiation into the IGM.

The partial covering of the emission by a DLA absorption system related  to the galaxy, together with the blue-shifted fan pattern, suggests that optically thick neutral hydrogen is falling in from the backside of the DLA host.  An extended screen appears to give rise to the spectroscopically amorphous fan, while a narrow, filamentary structure plausibly produces the spatially elongated but spectroscopically narrow ridges.  The latter component could represent a tidal tail of HI gas or a cold accretion filament connected to the galaxy.  Indeed, the observational properties of the narrow structure, including the thinness, the quiescent velocity field as evident in the emission ridges, the density estimate, and the total column density are well consistent with those predicted for such filaments (e.g., Keres et al 2005; Dekel et al 2009; Goerdt et al 2010; Faucher-Giguere et al 2010).  A possible tilt of the emission ridges, with their shifting further to the blue when approaching the continuum trace of the galaxy,  could be understood within this scheme as accelerated infall, although other explanations may be possible. If our interpretation is correct, this would be the first detection of cold accretion in an ordinary high redshift galaxy. The visibility of such a structure in this case is due to external illumination, leading to fluorescent Ly$\alpha$ emission.

The relatively simple velocity structure and the lack of spatial gradients in the Ly$\alpha$ surface brightness suggest that the Ly$\alpha$ emission outside of the red core arises when the external gas is exposed directly to stellar ionizing radiation escaping from the galaxy. We have shown that the stellar ionizing radiation is sufficient to explain the Ly$\alpha$ flux observed. We infer a significant ($\sim$50\%) escape fraction  for ionizing photons. Cooling radiation and galactic-scale shocks may also play a role in the production of additional Ly$\alpha$ photons, and the latter may boost the ionizing radiation field, but both processes are unlikely to dominate over stellar radiation.

We have drawn attention to the importance of galactic encounters for the escape of Ly$\alpha$ and ionizing radiation at high redshift. Tidal acceleration or gas pressure may create velocity fields with optically thin conditions for the escape of Ly$\alpha$ in velocity space (see Appendix B). Actual physical breaches in the gaseous halos, "naked" tidal tails  (partly stripped of neutral hydrogen) containing hot stars, or simply distended gaseous halos may allow for the escape of ionizing continuum. 

Given the apparent strong asymmetry in the stellar population and in the Ly$\alpha$ halo of our object,  it is likely that the emission of ionizing radiation is highly anisotropic. With the radiation leaving predominantly in the backward direction, the part of the gaseous halo facing the observer may remain intact and the object might not appear in direct searches for escaping Lyman continuum. This picture of asymmetric escape of ionizing radiation could  provide a physical explanation for the findings by Iwata et al (2009) and Nestor et al (2011), that there are spatial offsets between the putative Lyman continuum emission and the non-ionizing continuum of their galaxy candidates.

While this object is so far unique, it may be no less common than merger remnants at that redshift.  We see only one such object in our survey, though this is not surprising given the small survey volume.  With the exception of the somewhat deeper FORS/VLT survey by Rauch et al (2008), which, however, had only 43\% of the volume of the current LDSS3/Magellan survey, the low surface brightness extended emission features could not have been detected previously. The discovery of this object was apparently helped by the presence of in-falling gas that intercepted part of the escaping ionizing flux and converted it into fluorescent Ly$\alpha$. Without this screen, half of the ionizing photon flux would have escaped into empty space, and only the relatively weak red core emission would have been visible.

This object may be the first detection of a once relatively common population of low-luminosity, interacting high-redshift galaxies.  If the scenario we have proposed is correct, then the disruption of HI halos may be an important mechanism for the escape of both Ly$\alpha$ and ionizing continuum radiation.  The object may therefore demonstrate a plausible physical mechanism by which faint galaxies are able to make a significant contribution to the ionizing emissivity at high redshift, and perhaps even dominate the ionizing photon budget during reionization.  

\section*{Acknowledgments}

We acknowledge useful discussions with Hsiao-Wen Chen, Li-Zhi Fang, Pat McCarthy, Masami Ouchi, and Francois Schweizer. MR is grateful to the IoA in Cambridge for hospitality in August 2010, and MH thanks Caltech for hospitality as a Kingsley visitor in April 2011. GB has been supported by the Kavli Foundation.

\newpage

\appendix

\section[]{Fast radiative shocks as sources of HI Ly$\alpha$, HeII 1640, and ionizing radiation}

This section examines the circumstances in which shocks can produce Ly$\alpha$ and HeII 1640 from a galaxy like the  present one. We also estimate the strength of ionizing radiation to see whether some of the observed "escape" of ionizing photons from high redshift galaxies may possibly have been produced by shock ionization.

We use a model output from the MAPPINGS library (Allen et al 2008), which, for illustrative purposes, is their model T\_n0.01\_b0.001. For a density 0.01 cm$^{-3}$, and solar abundances, the expected Ly$\alpha$ luminosity  from shock and precursor for a shock with area A would be

\begin{eqnarray}
L_{{\rm Ly}\alpha} = (7.1,38,93,152)\times 10^{37}   \left(\frac{A}{{\rm kpc}^2}\right)  {{\rm erg}\ {\rm s}^{-1}},
\end{eqnarray}
corresponding to shock velocities of 200, 300, 400 and 500 km,s$^{-1}$, respectively. If the entire observed size of the Ly$\alpha$ emitter ($\sim 5"\times 2 "$, corresponding to 580 kpc$^2$) were undergoing such a shock the total luminosity would range up to $8.8\times 10^{41}$erg s$^{-1}$ (for shock velocity 500 km\,s$^{-1}$), within a factor of a  1/3 of the observed Ly$\alpha$ luminosity $2.46\times10^{42}$ erg s$^{-1}$. 

For the fiducial MAPPINGS models presented by Allen et al (2008), the HeII 1640/Ly$\alpha$ ratio for shock ionization is generally higher  than observed here. It reaches a minimum between $300< v_{sh}< 500$ km\,s$^{-1}$ near 0.13, however, which is not very different from the observed ratio of 1/10. Thus, plausible shock parameters can be found that lead to absolute HI Ly$\alpha$ luminosities and HeII 1640/Ly$\alpha$ ratios that are not altogether different from the observed ones. 

The interesting question remains whether a large scale shock could contribute significantly to the ionizing flux from a high redshift galaxy, and thus the  reionization photon budget, raising the apparent "escape fraction" for Lyman continuum radiation. 

For a galaxy at z=3.34, of magnitude $m_{AB}$, with an intrinsic continuum slope S=L(1500\AA)/L(900\AA), and a relative escape fraction $f_{esc}$, the escaping luminosity density at 900\AA\  would be
\begin{eqnarray}
L_{\nu}^{esc} = 2.4\times10^{26}\left(\frac{S}{6}\right)^{-1}\left(\frac{f_{esc}}{0.1}\right)10^{-0.4 (m_{AB}-27))}\frac{\rm erg}{\rm s\ Hz}.
\end{eqnarray}
For the aforementioned MAPPINGS models, the luminosity  density  at 900\AA, radiated into $4 \pi$  is
\begin{eqnarray}
L_{\nu} = 2.6(17,45,72)\times10^{22} \left(\frac{{\rm A}}{{\rm kpc}^2}\right) \frac{\rm erg}{\rm s\ Hz},
\end{eqnarray}
where $A$ again is the area of the shock in kpc$^2$. Assuming again that the entire Ly$\alpha$ emitting regions (size A=580 kpc$^2$) undergoes the shock, the resulting maximum luminosity density (for 500 km\,s$^{-1}$) rises to $4.2\times 10^{26}$ erg s$^{-1}$ Hz$^{-1}$,  close to an 18\% percent escape fraction for an $m_{AB}\sim 27$ galaxy, as discussed above. We note, however, that in our case the emulated escape fraction is so high only because it belongs to a rather faint galaxy for which we have assumed enormous shock velocities. This would therefore be an extreme case.  For brighter galaxies, where the escape fraction can be measured with broad band images, shocks may be far less relevant.

\section[]{Escape of L\lowercase{y}$\alpha$ (and ionizing radiation) in a tidal velocity field}

Several lines of evidence suggest that the Ly$\alpha$ emitter stage may be a transient phenomenon in the lifetime of a galaxy.  Earlier work has described tensions between the clustering of Ly$\alpha$ emitters and that of dark matter halos (e.g., Hayashino et al 2004; Hamana et al 2004; Kovac et al 2007; Zheng et al 2011), and a limited duty cycle for Ly$\alpha$ emission has been proposed to account for these discrepancies (Shimizu et al 2007). It may also explain observed number counts  of Ly$\alpha$ emitters  (Malhotra \& Rhoads 2002), and  reconcile the rates of incidence of Ly$\alpha$ emitters and damped Ly$\alpha$ systems (Barnes \& Haehnelt 2009). While the  properties of Ly$\alpha$ emitters have mostly been discussed (but not completely understood) in terms of their stellar populations, it may instead be the gas dynamics that decides whether a galaxy appears as a Ly$\alpha$ emitter, or not.

\begin{figure}
\includegraphics[scale=.31,angle=0,keepaspectratio = true]{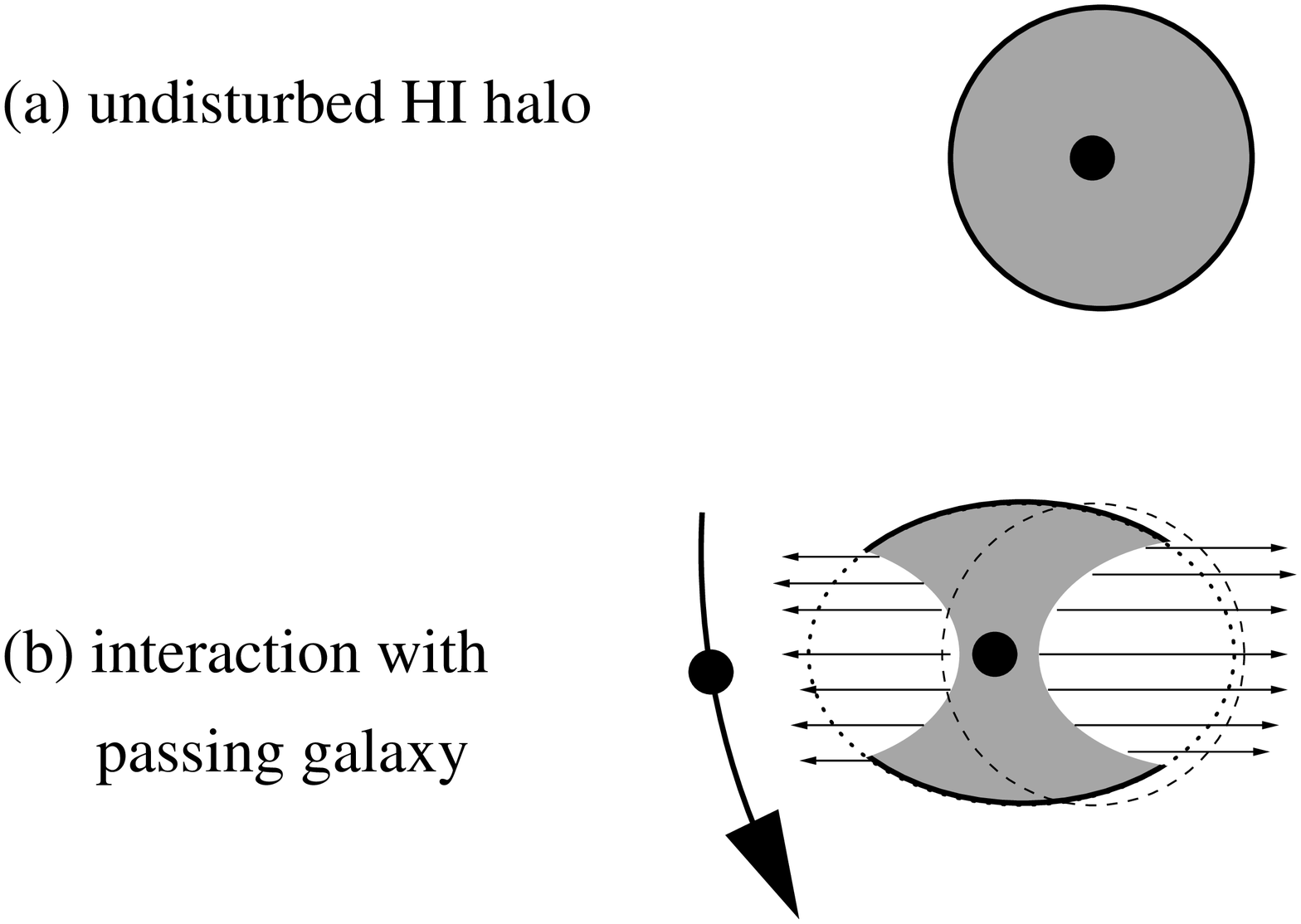}
\caption{Release of Ly$\alpha$ radiation by tidal (or ram pressure) interaction with a nearby galaxy. In this highly schematic depiction, the top panel shows an undisturbed HI halo (grey) surrounding a Ly$\alpha$ producing galaxy (dark spot). The Ly$\alpha$ photons escape gradually through the optically thick HI coccoon. In the bottom drawing, which is shown in the reference frame of the galaxy's center of mass, the encounter with a perturbing galaxy (e.g., in the form of a tidal field) leads to differential motion between the Ly$\alpha$ source and the gaseous halo in the part closest to the passing galaxy and at the opposite (far) end of the halo. If the relative velocities between gas and galaxy in a certain direction begin to exceed the thermal velocities, the photons drop out of the original resonance and may escape preferentially through these "velocity channels". The Ly$\alpha$ emission will brighten up in these high velocity directions and may boost the flux seen along this line of sight across a detection threshold. The lower optical depth and thus shorter path through the medium will also reduce the chances for absorption by dust. 
\label{fig4}}
\end{figure}

We briefly discuss here how interaction with another galaxy may induce a temporary increase in the escape of Ly$\alpha$ photons, allowing a galaxy to appear as a Ly$\alpha$ emitter.  The escape of Ly$\alpha$ may be facilitated by tidal acceleration, even in cases where the galaxies remain intact and the gaseous halos retain sufficient neutral hydrogen to be optically thick to ionizing radiation. Fig.(\ref{fig4}) shows how differential acceleration caused by the tidal force exerted by a passing galaxy may be able to  boost the Ly$\alpha$ flux escaping in the direction both toward the passing object and in the opposite direction. The near part of the gaseous halo is more strongly accelerated toward the passing galaxy than the central source of the Ly$\alpha$ photons, which in turn is more strongly accelerated than the gas on the distant side.  To first order, this will allow Ly$\alpha$ photons from the central source to drop out of resonance with the surrounding gas, and hence escape.  The resulting double-component structure may mimic the Ly$\alpha$ spectroscopic emission pattern of a bi-polar galactic outflow. The flare-up of the Ly$\alpha$ activity caused by the distorted halo would occur on a dynamical time-scale for the passage of the disturbing galaxy, and die off  when the gaseous halo relaxes over a free-fall time for the parent galaxy. 

This model may explain why Ly$\alpha$ emission is preferentially found in close pairs of Lyman break galaxies (Cooke et al 2010). It would also explain the apparent transience of Ly$\alpha$ through a dynamical mechanism, rather than relying on timescales directly proscribed by the lifetimes of a stellar population. If star formation also gets triggered by galactic interactions, then these processes, of course, will not be independent. 

And finally, when tidal forces distend the gaseous halo, they reduce the (column) density and optical depth for ionizing radiation, even if they fall short of actually exposing stellar populations directly, e.g., in tidal tails.  This process, which is a milder version of the case  discussed in this paper, would associate enhanced Lyman continuum escape preferentially with Ly$\alpha$ emitters, as apparently observed by Nestor et al (2011).

\bigskip

\bsp

\label{lastpage}

\end{document}